# Image-Plane Detection of Spatially Entangled Photon Pairs with a CMOS Camera

David McFadden[1,2,3] and Rainer Heintzmann[1,2,3]

[1] Institute of Physical Chemistry and Abbe Center of Photonics,
Friedrich-Schiller-University, Jena, Germany

[2] Leibniz Institute of Photonic Technology, Albert-Einstein-Straße 9, 07745 Jena, Germany

[3] Jena Center for Soft Matter (JCSM), Friedrich Schiller University Jena, Jena, Germany

Author for correspondence: David McFadden, e-mail: david.andrew.mcfadden787@gmail.com

## ABSTRACT

Spatially entangled photon pairs (biphotons) generated by spontaneous parametric down-conversion offer unique opportunities for quantum imaging, but image-plane biphoton correlations are difficult to observe with camera-based detectors. Previous camera-based biphoton imaging experiments have relied on photon-counting detection, which necessitates operation deep in the photon-sparse regime and requires extremely low dark rates.

Here, we demonstrate the detection of spatial biphoton joint probability distributions in both the image plane and the pupil plane (also termed "near-field plane" and "far-field plane" respectively) using a conventional scientific CMOS camera operated in linear mode. We work at mesoscopic intensity levels, corresponding to a photon flux approximately four orders of magnitude higher than typical photon-counting approaches. From the measured image- and pupil plane correlations, we observe position and momentum correlations consistent with an EPR-type entanglement witness.

A tailored correlation analysis suited for image plane imaging suppresses detector artifacts and intensity fluctuations, enabling acquisition with significantly fewer frames. Our results demonstrate that spatially entangled-light imaging can be performed efficiently with standard imaging hardware, extending quantum imaging techniques beyond the photon-counting regime.

**Introduction**

Imaging using light from spontaneous parametric down-conversion has attracted attention for use in imaging. This light consists of biphotons which are position-correlated as they impinge on the detector in the image plane. Images are recovered by detecting and registering these joint arrivals of two photons. This is in contrast to classical imaging, where all arriving photons contribute equally to the image.

Following early theoretical predictions, several experiments have demonstrated that joint detection of biphoton distributions can offer advantages in resolution, contrast, or other information content compared to classical illumination.[1,2] These experiments also highlight a number of experimental challenges. Biphoton light is fragile to optical loss and imaging conditions on the detector. A central experimental challenge lies in the detection of spatial correlations at the single-photon level. Conventional CCD and CMOS cameras are well suited for fast, high-resolution imaging, but their readout noise obscures correlations when operating in the single-photon regime. As a result, most biphoton imaging experiments rely on photon-counting detectors, including intensified CCDs, EMCCDs, and SPAD arrays.

Photon counting techniques eliminate readout noise, but impose a severe constraint on the usable photon flux. In image-plane biphoton imaging, accurate correlation measurements require the incident intensity to remain well below one photon per pixel per frame on average, ensuring that joint detection events can be unambiguously interpreted. This constraint leads to long acquisition times and typically very large frame counts.

Advances in active pixel sensor technology have led to scientific CMOS cameras with sub-electron readout noise, high quantum efficiency, and large pixel counts. While such cameras have replaced CCD and EMCCD detectors in many imaging applications, their use in biphoton imaging has largely remained restricted to photon-counting-like operation at very low light levels. [3]

Here, we demonstrate detection of spatial quantum correlation at photon fluxes of ~200 photons per pixel per frame, well in excess of the photon-sparse regime. While this has been pointed out by Defienne et al. [4] for pupil plane correlations, we are able to recover the spatial correlations in the image plane.

**Experimental setup**

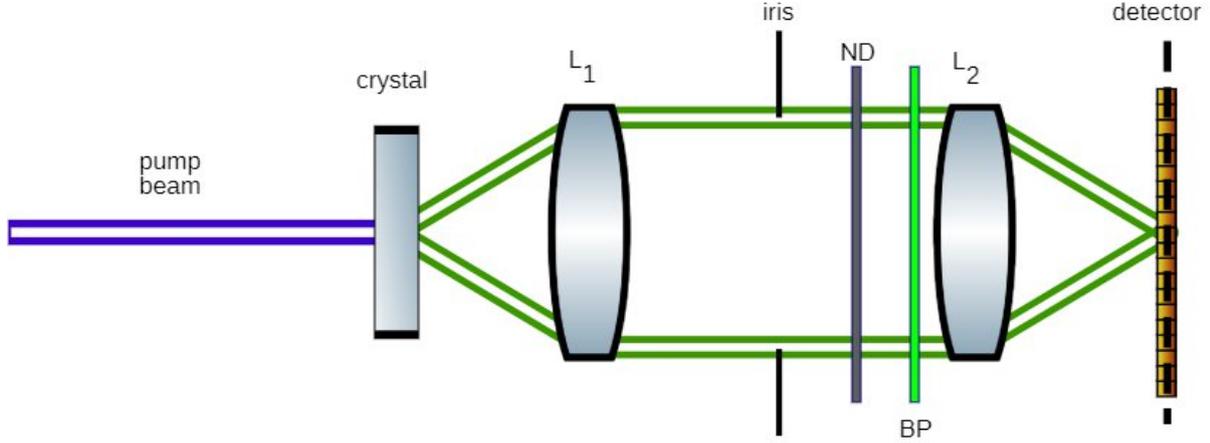

Figure 1: Schematic of the biphoton imaging setup.

The pump beam is provided by a continuous-wave laser at a wavelength of 266 nm and is incident on a β-barium borate (BBO) crystal cut for type-I spontaneous parametric down-conversion.

Following the crystal, a reflective long-pass filter suppresses the pump while transmitting the down-converted light. The SPDC field is then relayed using an infinite conjugate imaging configuration. A first lens ($L_1$) with a focal length of 150 mm is placed one focal length from the crystal. An adjustable iris is positioned one focal length behind lens $L_1$. This is the pupil plane or Fourier plane of the imaging system.

A second lens ($L_2$), with a focal length, of 250 mm is placed behind the iris and images the crystal plane onto the camera sensor, thereby completing the imaging system. This is the image plane configuration.

When imaging the pupil plane, the lens ($L_2$) is removed and replaced by an alternative relay lens system ($L_3$) with a focal length of 57 mm, placed in a position such that the iris plane is imaged onto the camera sensor without moving the sensor.

A band-pass (BP) filter with a central wavelength of 520 nm, close to the degenerate wavelength of the SPDC process, is placed in the optical path downstream of the iris, but before $L_2$ or $L_3$.

Biphoton correlations are fragile to loss, and we exploit this sensitivity as an experimental control: by introducing an additional neutral density filter (ND = 0.6), which blocks about 75 percent of the light, along with the bandpass filter, we significantly reduce biphoton pairs, while we can alternatively reduce the pump power to an identical detected photon flux maintaining the fraction of biphotons in the detected light but affecting the spurious correlations.

**Results**

We acquired 1500 frames (a stack) at a peak photon detection rate of 150 photons/pixel/frame in the image plane configuration. In the pupil plane configuration, we acquired 1500 frames at a peak photon detection rate of 130 photons/pixel/frame.

Both of these acquisitions were repeated with the neutral density filter (ND=0.6) inserted into the detection path and the power of the pump accordingly increased such that the photon flux at the detector was close to the previous experiments, reducing the biphoton arrival rate to approximately ($0.25^2$ = 6.25%). This represents our control experiment.

In the pupil plane, the constituent photons of an SPDC pair are expected to arrive at opposing positions on the detector due to them exhibiting momentum anticorrelation after being generated at the SPDC crystal. In contrast, in the image plane, the two photons of each biphoton pair are spatially correlated and arrive at close-by transverse positions.

To extract the various correlations, we first calculate the mean projection of each stack and subtract it from the individual frames, leaving us with the residual fluctuations about the mean value. For the pupil plane recordings we compute the 2D autoconvolution of each frame, and then average over the resulting individual anticorrelation maps. The result is an estimate of the 2D joint probability distribution for momentum anti-correlated detection events.

For the mean-subtracted image plane recordings, we compute the 2D autocorrelation of the individual frames and average the results. This results in a joint detection probability map for spatially correlated photons. We estimate a correlation background, which can be attributed to detector artifacts and nonlinearities, by correlating each frame $n$ with a subsequent fame $n + 2$. This averaged correlation background is then subtracted from the averaged autocorrelation image.

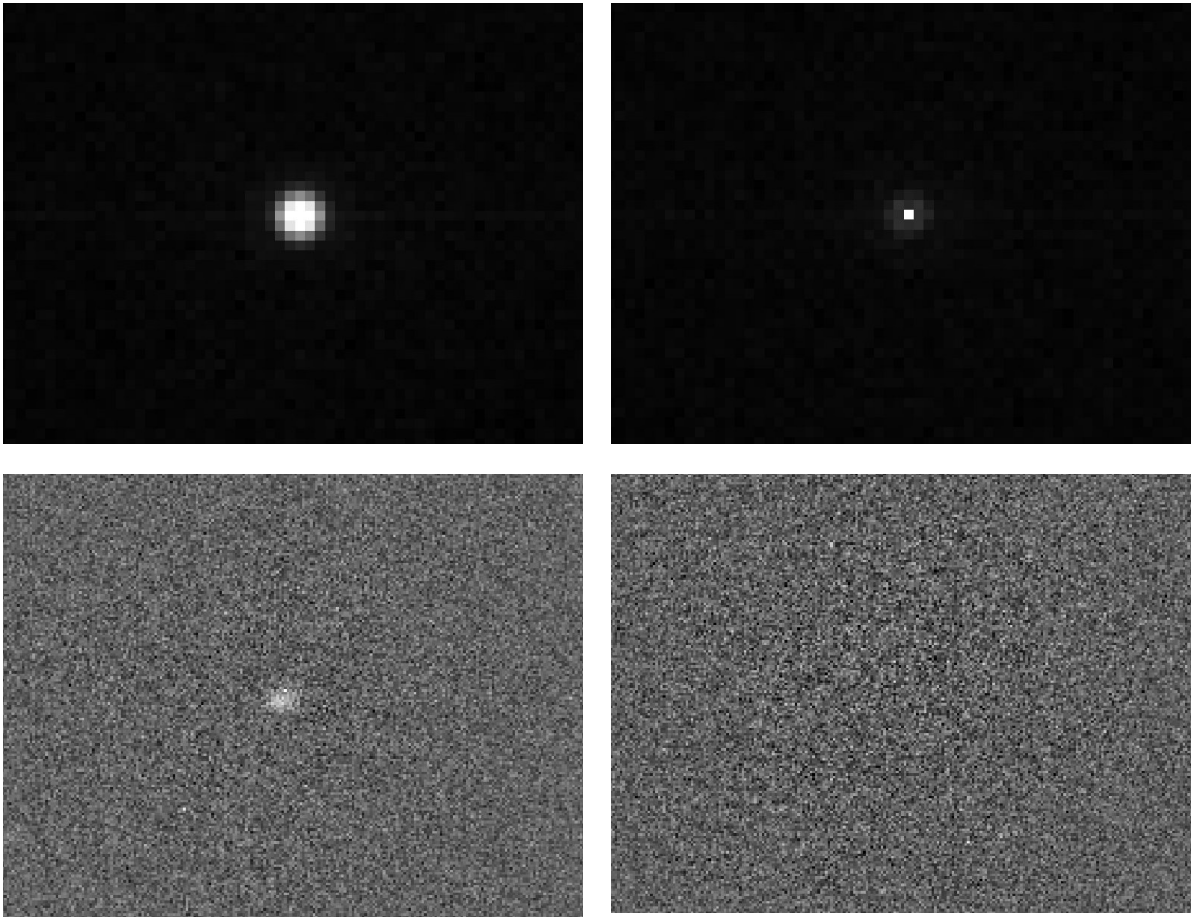

Figure 2: Correlation and anticorrelation maps. Top left: The central peak in the correlation image from acquisition in the image plane configuration. Top right: The same image region in the control experiment using downstream attenuation of the SPDC light. In both displays, the maximum brightness was set to the same value. Bottom left: A region of interest around the central peak in the anticorrelation image in the pupil plane configuration. Bottom right: The same region from the control experiment with attenuated light. Each pixel corresponds to 6.5μm × 6.5μm on the detector.

The resulting images are shown in figure 2. They exhibit peaks, which are greatly diminished in the control experiment with attenuated light.

An EPR paradox (EPR steering) is demonstrated when the product of inference variances violates the Heisenberg limit for the signal mode [5,6]:

$$\Delta^2_{min}(x_1|x_2)\Delta^2_{min}(p_{x1}|p_{x2}) < \frac{\hbar^2}{4}.$$

We estimate the widths of inference variances via a fit of a 2D Gaussian model optimizing for a minimal sum of square difference. As a discrete pixel is necessarily highly correlated with itself, we opt to exclude the central pixel from the fit for the image plane data. Our fit was performed over a circular central region 80 pixels in diameter.

The resulting estimates of variances from a least squares fitting are

$$\Delta^2_x = (6.97\ \mu m)^2$$
$$\Delta^2_p = (6394\ \hbar/m)^2$$

The product of these is $1.98 \times 10^{-3}\hbar^2$. This violates the inequality by a factor of more than two orders of magnitude.

**Influence of imaging parameters**

We also investigated the influence of the imaging parameters by repeating this experiment for different diameters of the iris aperture. Varying the iris diameter changes the effective numerical aperture of the imaging system and, consequently, modifies the size of the single-photon point spread function at the detector plane.

As the aperture is reduced, the transverse extent of the correlation peak broadens, while its height decreases. This behavior reflects the close relationship between the single-photon point spread function and the measured joint probability distribution: restricting the numerical aperture increases diffraction, causing detected photons to be distributed over a larger area and thereby reducing the local coincidence density.

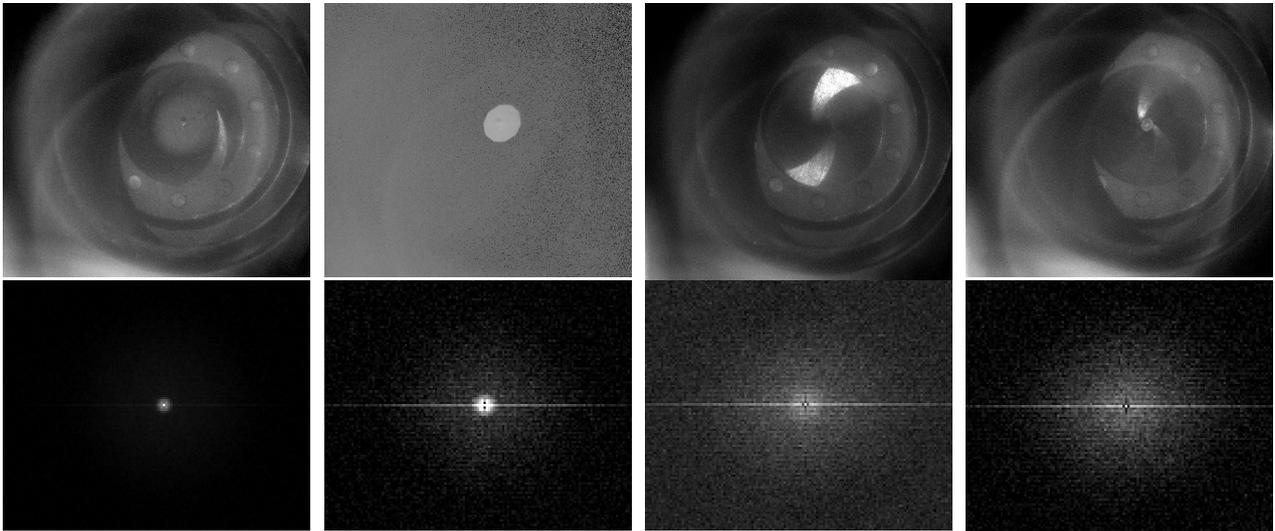

Figure 3: Top: Images of the pupil plane at aperture diameters of 12 mm, 6.2 mm, 2.4 mm, and 2.0 mm (visualization purposes only). Bottom: Correlation maps in the image plane corresponding to the aperture configurations shown above. The spatial width of the correlation peak increases with smaller apertures while its amplitude diminishes. The brightness levels have been adjusted for visibility. Each pixel corresponds to a correlation distance of 5μm.

## Discussion

As demonstrated above, we were able to analyze correlations both in the image plane as well as the anticorrelations in the pupil plane. We then measured the width of both of these correlations and compared the product with $\frac{\hbar^2}{4}$, according to Reid [7] to act as an EPR criterion. However, our findings are only to be taken as a hint of true entanglement in the data rather than an unambiguous proof. The reason is that our image processing removes spurious correlations, which would invalidate our variance estimation since they are present anywhere. The estimation of variance from a least-squares fit of a Gaussian profile is also not rigorous: Since our approximately round pupil has a hard-edged aperture, the PSF is expected to exhibit an infinite variance since the second moment of the PSF is proportional to the second derivative of the intensity OTF at zero spatial frequency, which is (at low NA) the autocorrelation of the pupil. The hard edge leads to an infinite curvature of the OTF at the center and thus an infinite variance. This clearly shows that our variance estimation by a least-squares fit of a Gaussian is unsuitable. Yet, our experiments still give a rough estimate of what would have been obtained if the pupil was apodized appropriately (e.g. Gaussian-shaped beams) leading to finite variances. A further point to note is that the image and pupil plane measurements were performed sequentially rather than simultaneously, which in principle leaves open a classical loophole.

## Conclusions

We have demonstrated the measurement of spatial biphoton joint probability distributions in both the image and pupil plane using a scientific CMOS camera. From these measurements, we observe position and momentum correlations consistent with an EPR-type entanglement witness, confirming the presence of spatial entanglement without relying on photon-counting detection.

By operating the detector in a mesoscopic regime with ~200 photons per pixel per frame, our approach avoids the severe flux limitations inherent to photon-counting techniques while remaining fully multimode.

These results extend the applicability of spatial quantum imaging beyond the photon-counting regime and open the door to faster, scalable investigations of entangled-light imaging and provide a practical route toward quantitative benchmarking of biphoton imaging techniques.